\documentclass[a4paper]{article}
\usepackage{verbatim}
\usepackage[left]{lineno}
\usepackage{amsfonts}
\usepackage{authblk} 
\newtheorem{theorem}{Theorem}

\newtheorem{corollary}[theorem]{Corollary}

\title{Faster construction of asymptotically good unit-cost 
error correcting codes in the RAM model}
\author{Djamal Belazzougui} 
\affil{Helsinki Institute for Information Technology (HIIT),
Department of Computer Science, University of Helsinki, Finland.}
\begin{document}
\maketitle
\begin{abstract}
Assuming we are in a Word-RAM model with word size $w$, 
we show that we can construct in $o(w)$ time an error correcting code
with a constant relative positive distance that maps 
numbers of $w$ bits into $\Theta(w)$-bit numbers, 
and such that the application of the error-correcting code 
on any given number $x\in[0,2^w-1]$ takes constant 
time. Our result improves on a previously proposed 
error-correcting code with the same properties 
whose construction time was exponential in $w$. 
\end{abstract}

\section{Introduction}
We work in the word-RAM model with word-size $w$. We assume  
that standard operations including multiplications (but not divisions) 
are supported in constant time. 
We present a way to construct an error correction over $O(w)$-bit strings
with the following features: 
\begin{itemize}
\item The code has some positive relative distance $\delta>0$. 
\item The evaluation of the code over any word takes constant time. 
\item The code can be constructed in time $o(w)$. 
\end{itemize}
Previously Miltersen~\cite{miltersen1998error} presented a code with similar
features except for the construction time which was exponential in $w$.
In the following, we denote by $H(x,y)$, the hamming distance between 
the two bitstrings $x$ and $y$. 
In what follows, we use the notation $[t]$ to denote the set 
$[0..t-1]$. We will often represent an integer $y$ of length $b$
as bitstring of length $b$ that consists in the concatenation 
of the $b$ bits of the number starting from most significant 
bit and ending in the least significant. 
We assume that $w$ is bigger than a sufficiently large constant. 
\section{The method}
Our method relies on code concatenation, a well known strategy in the design of 
error correcting codes. We will use the same error correcting code used by Miltersen~\cite{miltersen1998error}
Combined with a Reed-Solomon code~\cite{reed1960polynomial}. 

Our strategy is to cut the original key into pieces of $B=\lceil\log w\rceil$ bits each. 
We view a key $x$ of length $w$ as the concatenation of $\lceil\frac{w}{B}\rceil$ 
keys of length $B$ bits each. That is $w=b_1b_2\ldots b_r$ with $r=\lceil\frac{w}{B}\rceil$. 
We will form $5$ different numbers $x_1,x_2\ldots x_5$ as follows: 
$x_i=b_i0^{4B}b_{i+5}0^{4B}\ldots$ for all $i\in[1..5]$. The numbers can easily be formed through 
bit shifts and masking. 

We then multiply every $x_i$ by a suitably chosen number $z_r$ that will in fact
represent the generator polynomial of a Reed Solomon code of block length
$P$, where $P$ is a prime number between $2^B$ and $2^{B+1}-1$.
Such a prime can easily be determine in time $O(w^{0.525}\cdot\mathrm{polylog}(w))$
as follows. By the result of~\cite{baker2001difference}, it is well known 
that for sufficiently large $x$, there exists at least one prime between $x$ and $x+O(x^{0.525})$. 
One can thus find a prime between $2^B$ and $2^B+2^{0.525\cdot B}$ in 
$O(2^{0.525\cdot B}\cdot\mathtt{poly}(B))=O(w^{0.525}\cdot\mathrm{polylog}(w))$ time,
by using the deterministic primality test of~\cite{agrawal2004primes}.

The generator polynomial is $g(\gamma)=(\gamma-\alpha)(\gamma-\alpha^2)\ldots (\gamma-\alpha^r)$, 
where $\alpha$ is a generator for the finite field modulo $P$, 
$r=\lceil\frac{w}{5B}\rceil$ and all the numbers $\alpha^i$ are taken modulo $P$. 
More precisely $\alpha$ is a primitive root of unity of order $P-1$. That is $\alpha$
is such that $\alpha^{P-1}\equiv 1\pmod P$ and $\alpha^r\not\equiv 1\pmod P$ for 
all $r\in[1,P-2]$.  Such an $\alpha$ can be found in time $O(w^{1/4+\epsilon})$~\cite{shparlinski1996finding}.
The final representation of the polynomial will be a word $z=c_10^{4B-1}c_20^{4B-1}\ldots$, 
where $c_0,c_1\ldots$ are the coefficients of the polynomial. 
The construction can easily be done in $O(w/\log w)$ time as follows. We 
start with the word $z_1=c_10^{4B-1}c_20^{4B-1}c_30^{4B-1}\ldots$, where 
every $c_i$ are numbers of $B+1$ bits and $c_1=-\alpha$, $c_2=1$ and $c_i=0$ 
for all $i>2$. 
This is the representation of the monomial $(\gamma-\alpha)$. 
We then can induce the representation of the polynomial $(\gamma-\alpha)(\gamma-\alpha^2)$, 
by multiplying $z_1$ by the number $c_10^{4B-1}c_20^{4B-1}\ldots$, 
where $c_1\equiv-\alpha^2\pmod P$ and $c_2=1$. This results in a number $z'_2$
that contains $c'_10^{3B-2}c'_20^{3B-2}c'_30^{3B-2}\ldots$. We then need 
to execute the modulo $P$ operation on each of $c'_1$,$c'_2$ and $c'_3 $. 
This can easily be done if we had the division operation available. 
It is well known that division by a constant can be simulated by one multiplication by
a constant and bit shifts~\cite[16, p. 509]{knuth1973art}. 
The constant used in the multiplication can easily be computed
in $O(\log w)$ time. 
Now we execute the operations in parallel on the word $z'_2$, resulting 
in a word $z_2$ that contains $c_10^{4B-1}c_20^{4B-1}c_40^{3B-1}\ldots$, 
where $c_i=c'_i\bmod P$. 
We continue in the same way by multiplying by the representations of 
the monomials $(\gamma-\alpha^i)$, for all $i\in[3,r]$, until we get the number $z_r$,
the representation of the polynomial $g(\gamma)$. 
Note we can deduce the numbers $\alpha^1,\alpha^2\ldots
\alpha^r$ in total $O(r)$ time, by simulating the modulo
operation in constant time. 

We denote the result of the multiplication of the $z_r$ 
by a number $x_i$ followed by the parallel modulo $P$ operation 
by $f_1(x_i)$.

\subsection{Inner code}
Our inner code is constructed following the strategy of Miltersen~\cite{miltersen1998error}. 
We will use an exhaustive search to find a good multiplier $m$
that gives a good error correcting code for numbers from 
$[2^{B+1}]$ into $[2^{4(B+1)}]$. 
We will have to test at most $O(w^3)=w^{O(1)}$ different multipliers. 
Testing every multiplier will take time $O(w^2)$ (it can be improved 
to $O(w\log w)$ time by using bit-parallelism). 
Basically, we need to check for every pair of numbers $a,b$ 
whether $H(f(a),f(b))\geq\delta B$ for some suitably chosen $\delta$. 
Thus the total time will be $w^4\log w=w^{O(1)}$ in the worst case. 
We now define the function $f_2(x)$ as the multiplication of 
$x$ by the number $m$. It is clear that if $x=c_10^{4B-1}c_20^{4B-1}\ldots$, 
then the result will be the number $y=c'_10^{B-2}c'_20^{B-2}\ldots$, 
with $c'_i=c_i\cdot m$. 

\subsection{Final result}
Given a word $x$, we first build the $5$ words $x_i$
for $i\in[1,5]$. We then apply the Reed Solomon code on
each of them, resulting in $5$ numbers $y'_i=f_1(x_i)$ 
for $i\in[1,5]$, where each number is of length $2w$ bits. 
We then compute the numbers $y_i=f_2(y'_i)$.
The final result will be the concatenation of the numbers 
$y_1,\ldots y_5$ which is of length $10$ words. 

\subsection{Analysis}
It can easily be seen that the resulting code has a 
positive relative distance $\delta'>0$. 
Assume we have two numbers $a$ and $y$, 
decomposed as $x_1\ldots x_5$ and $y_1\ldots y_5$. 
Then if $x_i\neq y_i$ for any $i$, we will be sure that
$f_1(x_i)=$ will differ from $f_1(y_i)$ in at least 
$r+1=\lceil\frac{w}{5B}\rceil+1$ fields. 
Further $f_2(f_1(x))$ and $f_2(f_1(y))$ will 
differ in at least $(r+1)(\delta B)=(\lceil\frac{w}{5B}\rceil+1)\delta B$ bits
which is at least $\delta'w=(\delta w/5)$.

\subsection{Further reduction}
We can further improve the total preprocessing time 
to $o(w)$, recursing once more. 
That is, first finding a concatenation 
of two Reed-Solomon-codes, one over $\Theta(w)$ 
bits and the other on $\log w$ bits and 
concatenate the result with a good multiplier code 
over $\log w$ bit-numbers that can be found in 
time $O(\log^4 w\log\log w)=o(w)$. 
The total construction time will be dominated 
by the time to construct the Reed-Solomon code 
over $w$-bit numbers which will take $O(w/\log w)$ time. 

The end result is an ECC whose final output is doubled 
compared to the one shown in previous section.  
The final output will be of length $20$ words. 

We thus have proved the following theorem. 
\begin{theorem}
Assuming we are in a Word-RAM model with word size $w$, 
we can construct in $o(w)$ time, an error correcting code
with some relative positive distance $\delta>0$ and that maps 
numbers of $w$ bits into number of $20w$ bits 
and such that the application of the error-correcting code 
on any given number $x\in[2^w]$ can be done in time constant 
time. The description of the error correcting code occupies 
$O(w)$ bits of space. 
\end{theorem}

\section{Applications}
In~\cite{hagerup2001deterministic} it is shown 
how given a set $S\subset [2^w]$ with $|S|=n$, 
one can construct a hash function $f$ from $[2^w]$ into 
$[n^c]$ bits for some constant $c>2$ such that:
\begin{enumerate}
\item The hash function is injective on the set $S$. 
That is $|f(S)|=n$. 
\item The hash function can be constructed in time $O(n\log n)$
assuming the availability of some constants that depend only on 
$w$ and that can be computed in time exponential in $w$. 
\end{enumerate}
The algorithm uses as a component a unit-cost error correcting code from $w$
bits into $4w$ bits with positive relative distance. The error correcting 
code consisted in a single multiplication by a constant of length $3w$ bits. 
In the Word-RAM model, an algorithm is said to be weakly non uniform if it uses 
some precomputed constants that depend only on $w$. 
In the construction of~\cite{hagerup2001deterministic}, 
there are two sources of weak non uniformity. The first one is
due to the use of a constant needed for the error correcting code 
and the other one due to constants used in a procedure that computes the most significant 
bit in words in constant time. It turns out that the 
computation of the constants needed for the last operation can be done in $O(w)$ time. 
The computation of the constant needed for the error correcting code was the bottleneck, since it was not known how 
to compute the constants in better than time $2^{O(w)}$. With our construction, 
this is no longer a bottleneck, since we have shown that 
we can construct a suitable error correcting code in time $O(w)$. 
By plugging our error correcting code in place of the previous
one, the signature hash function can now be built in time $O(n\log n)$ whenever $w\leq n$, 
even when the time to compute the constants it taken into account. 
We thus have the following corollary: 
\begin{corollary}
Assuming we work in the Word-RAM model with word length $w$,
given a set $S\subset [2^w]$ with $|S|=n\geq w$, 
we can in $O(n\log n)$ time 
build a function that maps $S$ into the set $[n^{O(1)}]$. 
The function can be described in $O(w)$ bits of space. 
\end{corollary}

There exists a alternative signature function~\cite{ruvzic2009making}
that does not need precomputed constants that are costly to compute 
and that runs in time $\omega(n\log n)$ 
for certain word sizes (more precisely, in time $O(n+n\frac{\log^3n}{w}(\log\frac{w}{\log n})^3)$). 
Choosing $w=\log^{1+\epsilon} n$ for some $\epsilon>0$,
implies construction time $\Omega(n\log^{2-\epsilon}n)$. 
Thus the signature functions of~\cite{hagerup2001deterministic} have the fastest 
construction time depending only on $n$.
\bibliographystyle{plain}
\bibliography{unit_cost_ECC.bib}

\begin{thebibliography}{1}

\bibitem{agrawal2004primes}
Manindra Agrawal, Neeraj Kayal, and Nitin Saxena.
\newblock Primes is in p.
\newblock {\em Annals of mathematics}, pages 781--793, 2004.

\bibitem{baker2001difference}
Roger~C Baker, Glyn Harman, and J{\'a}nos Pintz.
\newblock The difference between consecutive primes, ii.
\newblock {\em Proceedings of the London Mathematical Society},
  83(03):532--562, 2001.

\bibitem{hagerup2001deterministic}
Torben Hagerup, Peter~Bro Miltersen, and Rasmus Pagh.
\newblock Deterministic dictionaries.
\newblock {\em Journal of Algorithms}, 41(1):69--85, 2001.

\bibitem{knuth1973art}
Donald~E Knuth.
\newblock {\em The Art of Computer Programming (Volume 3)}.
\newblock Addison--Wesley, 1973.

\bibitem{miltersen1998error}
Peter~Bro Miltersen.
\newblock Error correcting codes, perfect hashing circuits, and deterministic
  dynamic dictionaries.
\newblock In {\em Proceedings of the ninth annual ACM-SIAM symposium on
  Discrete algorithms}, pages 556--563. Society for Industrial and Applied
  Mathematics, 1998.

\bibitem{reed1960polynomial}
Irving~S Reed and Gustave Solomon.
\newblock Polynomial codes over certain finite fields.
\newblock {\em Journal of the Society for Industrial \& Applied Mathematics},
  8(2):300--304, 1960.

\bibitem{ruvzic2009making}
Milan Ru{\v{z}}i{\'c}.
\newblock Making deterministic signatures quickly.
\newblock {\em ACM Transactions on Algorithms (TALG)}, 5(3):26, 2009.

\bibitem{shparlinski1996finding}
Igor Shparlinski.
\newblock On finding primitive roots in finite fields.
\newblock {\em Theoretical computer science}, 157(2):273--275, 1996.

\end{thebibliography}

\end{document}